\documentclass[useAMS,usenatbib]{mn2e}
\usepackage{graphicx,graphics,amsmath}
\usepackage{latexsym}

\begin{document}
\def\ter{\theta_{\rm er}}
\def\si{\sin \theta}
\def\Ro{R_{\odot}}
\def\co{\cos \theta}
\def\diff{$\pa \Omega/\pa r$}
\newcommand{\Bf}{{\bf B}}
\newcommand{\ep}{{\bf e}_\phi}
\newcommand{\vf}{{\bf v}}
\newcommand{\pa}{\partial}
\newcommand{\Rs}{R_{\odot}}
\newcommand{\er}{\mbox{erf}}
\newcommand{\Bc}{B_c}
\def\Rs{R_{\odot}}
\def\mc{meridional circulation}

\title[The Waldmeier effect and the flux transport solar dynamo]{The Waldmeier effect and the flux transport solar dynamo}
\author[B.\ B.\ Karak and A.\ R.\ Choudhuri]
{Bidya Binay Karak $^{1}$ and Arnab Rai Choudhuri$^{2}$ \\
  Department of Physics, Indian Institute of Science, Bangalore 560012, India\\
  $^{1}$bidya$\_$karak@physics.iisc.ernet.in\\
  $^{2}$arnab@physics.iisc.ernet.in}
\date{}  
\maketitle
\begin{abstract}
We confirm that the evidence for the Waldmeier 
effect WE1 (the anti-correlation between rise
times of sunspot cycles and their strengths) and the related 
effect WE2 (the correlation between rise rates of cycles and their strengths)
is found in different kinds of sunspot data. We explore whether
these effects can be explained theoretically on the basis of
the flux transport dynamo models of sunspot cycles. Two sources of
irregularities of sunspot cycles are included in our model:
fluctuations in the poloidal field generation process and
fluctuations in the meridional circulation. We find WE2 to be
a robust result which is produced in different kinds of theoretical
models for different sources of irregularities.  The Waldmeier
effect WE1, on the other hand, arises from fluctuations in the
meridional circulation and is found only in the theoretical models
with reasonably high turbulent diffusivity which ensures that
the diffusion time is not more than a few years. 

\end{abstract}

\section{Introduction}

Waldmeier (1935) noted an anti-correlation between the rise times
of sunspot cycles and their strengths.  In other words, a cycle with
a longer rise time is expected to have a weaker peak at the maximum.
This is known as the Waldmeier effect.  We shall refer to this as
WE1.  There is another related effect.  The rise rates of cycles show
a correlation with their strengths: a faster rising cycle is likely
to be stronger.  We shall call it WE2.  Occasionally one uses the
term Waldmeier effect to also mean this second effect WE2, causing some
amount of confusion in the literature. For example, sometimes one
talks of using the Waldmeier effect to predict the strength of
a sunspot cycle after it has just begun. In this case, clearly WE2
which involves rise rates is meant rather than WE1
which involves rise times.  Shortly after a sunspot cycle has begun,
it becomes possible to estimate its rise rate, but it is not possible
to know the rise time until the cycle has reached its maximum.

The aim of this paper is to explore whether the Waldmeier effect
can be explained with the flux transport dynamo model, which presently
appears to be the most promising model for explaining the solar cycle.
The flux transport dynamo model involves several parameters, some of
which are rather poorly known at the present time. One important
question is whether the Waldmeier effect is reproduced theoretically
only for certain combinations of parameters.  If that is the case, then
it should be possible to put some constraints on the parameters 
of the flux transport dynamo by
demanding that the theoretical model accounts for the Waldmeier
effect.  We also present a short discussion of the observational
data, in view of a recent controversy whether the Waldmeier effect
really exists.  Hathaway, Wilson \& Reichmann (2002) found evidence for WE1 in both
the Z\"urich sunspot numbers and the group sunspot numbers.  But Dikpati, 
Gilman \& de Toma (2008) claim that this effect does not exist in sunspot area
data.  We argue that the rise time has to be properly defined to obtain
the Waldmeier effect.  In our opinion, Dikpati, Gilman \& de Toma (2008) failed
to discover WE1 in the sunspot area data because they had not taken
proper rise times.  With a proper definition of the rise time, we show
that WE1 is present in various kinds of sunspot data.  The other effect
WE2 seems more robust.  Cameron \& Sch\"ussler (2008) found evidence
for WE2 in various kinds of sunspot data, which we also confirm.  Thus,
in our view, both WE1 and WE2 are real effects which a satisfactory
theoretical model of the sunspot cycle should explain.

Let us mention some of the salient features of the flux transport dynamo
model, which has been developed by many authors during the last few
years (Choudhuri, Sch\"ussler \& Dikpati 1995; Durney 1995; Dikpati \& Charbonneau 1999;
K\"uker, R\"udiger \& Schultz 2001; Nandy \& Choudhuri 2002; Choudhuri 2003; 
Chatterjee, Nandy \& Choudhuri 2004; Choudhuri, Chatterjee, \& Nandy 2004; 
Mu\~noz-Jaramillo, Nandy \& Martens 2009). The toroidal magnetic field is produced in the
tachocline by the action of differential rotation on the poloidal field
and eventually rises to the solar surface due to magnetic buoyancy to
produce sunspots.  The decay of tilted bipolar sunspots gives rise to a
poloidal field near the surface by the mechanism first elucidated by 
Babcock (1961) and Leighton (1969). The meridional circulation, which is
observed to be poleward in the upper half of the solar convection zone
(SCZ) and must have a hitherto unobserved equatorward component at the
bottom of SCZ to conserve mass, advects the toroidal field equatorward
at the bottom of the SCZ and advects the poloidal field poleward at the
surface.  This provides the theoretical explanation of the equatorward
drift of sunspot belts as well as the poleward migration of the weak diffuse
magnetic field on the solar surface.  Lastly, we need a mechanism to transport
the poloidal field from the surface where it is generated by the
Babcock--Leighton mechanism to the bottom of SCZ where differential
rotation can act on it.  This transport of the poloidal field can be achieved
by two means: through advection by the meridional circulation or through
diffusion.  Yeates, Nandy \& Mackay (2008) have divided flux transport dynamo
models into two classes: advection-dominated and diffusion-dominated, 
depending on the transport mechanism of the poloidal field from the
surface to the bottom of SCZ.  Jiang, Chatterjee \& Choudhuri (2007) and Yeates, Nandy \& Mackay (2008)
were the first to point out many qualitative differences between these
two kinds of models.  Many authors (Chatterjee, Nandy \& Choudhuri 2004; Chatterjee
\& Choudhuri 2006; Jiang, Chatterjee \& Choudhuri 2007; Goel \& Choudhuri 2009; Choudhuri
\& Karak 2009; Hotta \& Yokoyama 2010a, 2010b) have given several independent 
arguments that the solar dynamo is likely to be diffusion-dominated.
We shall show in this paper that only diffusion-dominated dynamos and
not advection-dominated dynamos can account for the Waldmeier effect WE1,
further strengthening the case that the solar dynamo is diffusion-dominated.
 
The readers should be cautioned that in the early years of flux transport
dynamo research sometimes the term `advection-dominated' was used rather
loosely and may not always conform with our present usage.  In this paper,
we shall use the terms `advection-dominated' and `diffusion-dominated' following
the careful classification introduced by Yeates, Nandy \& Mackay (2008; see their
Fig.\ 7a). It should also be noted that at the bottom of SCZ the advection
of the toroidal field by the equatorward meridional circulation has to be
the dominant process over diffusion, as emphasized by Choudhuri, Sch\"ussler \& Dikpati
(1995).  If this were not the case, then the dynamo wave would propagate
poleward, following the dynamo sign rule (Parker 1955; Yoshimura 1975;
Choudhuri, Sch\"ussler \& Dikpati 1995; see also  Choudhuri 1998, 
\S16.6). To ensure this dominance of advection at the bottom of
SCZ, the diffusion has to be very low in the tachocline.  However, the
diffusion can be much larger within the convection zone to make the
transport of poloidal field across the SCZ diffusion-dominated.

In order to explain the Waldmeier effect, we need to understand what makes
the sunspot cycles unequal.  In the flux transport dynamo models, the
period of the cycle roughly scales as the inverse of the meridional circulation
amplitude.  Different authors have reported scaling 
laws with the power law indices fairly
close to $-1$: Dikpati \& Charbonneau (1999) reporting an index of $-0.89$
and Yeates, Nandy \& Mackay (2008) reporting $-0.885$.  Fluctuations in the meridional
circulation are expected to make the cycles unequal---making some longer and
some shorter.  We discuss our present knowledge (or lack of knowledge) of
meridional circulation fluctuations in \S2 and then introduce these fluctuations
in our theoretical calculations.  The other main source of irregularities in
the solar dynamo is the fluctuations in the Babcock--Leighton process, which  
involves decay of tilted bipolar regions.  Since this tilt is produced by
the Coriolis force acting on the rising flux tubes (Choudhuri 1989; D'Silva \& Choudhuri 1993)
and the rising flux tubes are buffeted by convective turbulence during
their rise, we expect a scatter in the tilt angles (Longcope \& Choudhuri 2002),
introducing a randomness in the Babcock--Leighton process. Choudhuri, 
Chatterjee \& Jiang (2007) identified this as the main source of randomness in the solar
dynamo.  They argued that the cumulative effect of these fluctuations is that
the poloidal field generated at the end of a cycle differs from the average
obtained in a mean field model.  According to Choudhuri, Chatterjee \& Jiang (2007), the
essential physics can be captured by multiplying the poloidal field above
$0.8 \Rs$ at the end of a cycle by a number $\gamma$ having random values in
a range around 1. The poloidal fields produced in earlier cycles are 
expected to be below $0.8 \Rs$ and are not affected.
We follow this procedure in this paper to study the
effect of fluctuations in the Babcock--Leighton process.

We check whether the effects WE1 and WE2 are produced in our theoretical
models on introducing irregularities due to fluctuations in the Babcock--Leighton
process and fluctuations in the meridional circulation. When the meridional
circulation (which determines the period in the flux transport dynamo) is held fixed,
fluctuations in the Babcock--Leighton process make the strengths of the different
cycles unequal, without varying the durations of the cycles or rise times too 
much---especially if the dynamo is diffusion-dominated, as we shall see.
Hence WE2 is the main effect which is relevant in this situation and not WE1.
We find that both diffusion-dominated and advection-dominated dynamos show
WE2 in this situation.  On the other hand, fluctuations in the meridional
circulation make the durations of cycles as well as rise times unequal, and one
can look for both the effects WE1 and WE2 in theoretical models on introducing
such fluctuations.  We find that only the diffusion-dominated model gives rise
to these two effects and not the advection-dominated model.  The physical reason
behind this remarkable result can be given on the basis of the analysis of
Yeates, Nandy \& Mackay (2008), as we shall point out in the appropriate place.

\section{Inputs from observational data}

We take a brief look at the sunspot cycle data to confirm that the Waldmeier
effect really exists and also discuss what we can say about fluctuations of meridional
circulation at the present time.

\subsection{Confirmation of the Waldmeier effect}

We have studied four different data sets: (1) Wolf sunspot numbers
(cycles 12--23), (2) group sunspot numbers (cycles 12--23), (3)
sunspot area data (cycles 12--23) and (4) $10.7$~cm radio flux
(available only for the last 5 cycles).  All data sets have been
smoothed by a Gaussian filter with a FWHM of $1$~yr.  

\begin{figure*}
\centering
\includegraphics[width=1.00\textwidth]{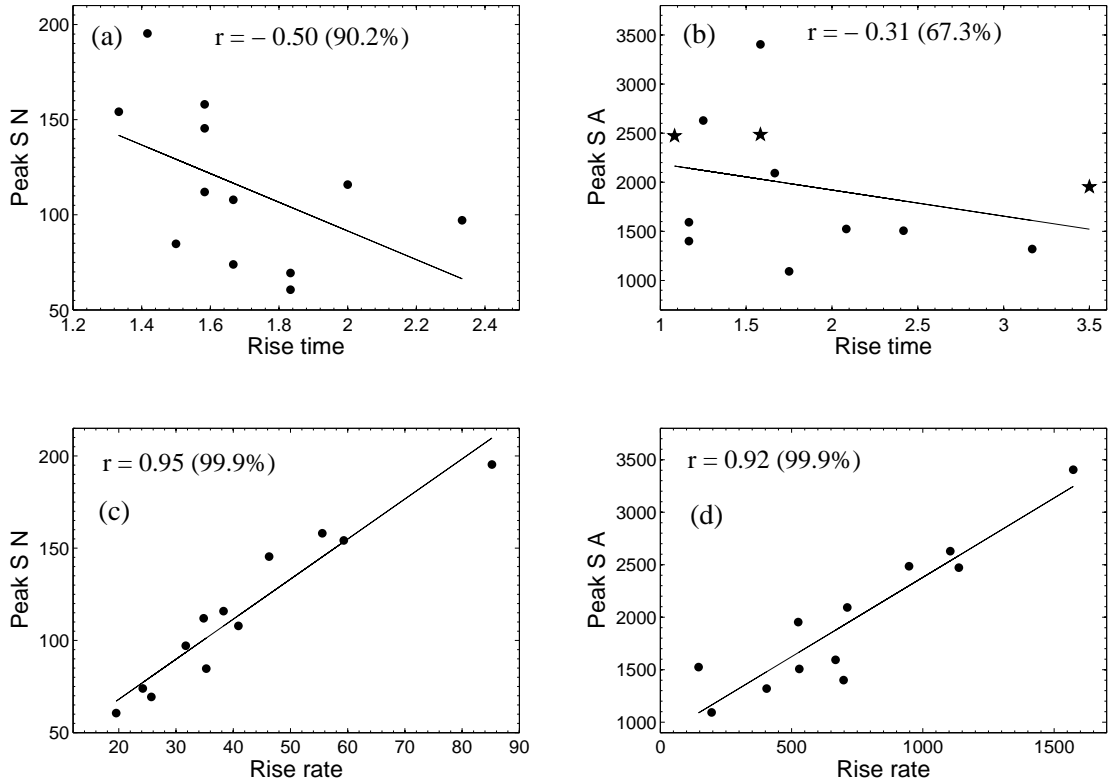}
\caption{Observational evidences for WE1 (upper row) and WE2 (lower row). The
upper row shows scatter diagrams plotting the peak values of (a)
sunspot number and (b) sunspot area against rise times (in years).
The data points of cycles 21, 22 and 23 are shown by stars in (b). The lower row
shows scatter diagrams plotting the peak values of (c) sunspot number
and (d) sunspot area against rise rates (in units of sunspot number per year).
The straight lines are the best linear fits of the data. The linear
correlation coefficients and the significance levels are also given on each plot.}
\label{period}
\end{figure*}

If the rise time is taken as the time for a cycle to develop
from a minimum to a maximum, then we have various difficulties.
Usually we find an overlap between two successive cycles and
the position of the minimum may get shifted depending on the
amount of the overlap (Cameron \& Sch\"ussler 2007). Some cycles have
plateau-like maxima with multiple peaks, so that it is difficult to
say when the rising phase ended. If one of the later peaks is
slightly higher than the earlier peak and one takes the later
peak to indicate the end of the rise time, then one gets a
``rise time'' without any physical significance.
This problem can be clearly seen
in Fig.~1 of Dikpati, Gilman \& de Toma (2008), where peaks in sunspot
area for the cycles 21, 22 and 23 are seen to occur well after
the cycles have reached the plateau-like tops.  We see in
the correlation plot given in Fig.~2 of Dikpati, Gilman \& de Toma (2008) 
that these are some of the cycles which produced the maximum
scatter and made the correlation disappear.  To avoid the 
difficulties of ascertaining the
minima and the maxima of the cycles, we define
the rise time in the following way.  Suppose a cycle has an
amplitude $P$. We take the rise time to be the time during which the
activity level changes from $0.2 P$ to $0.8 P$.  The rise time defined
in this way has a good anti-correlation with the cycle amplitude for
all the data sets, the linear correlation coefficients and the significance
levels for the four data types being: (1) $-0.50$ and $90.16\%$ for
sunspot numbers; (2) $-0.42$ and $82.12\% $ for group sunspots; (3)
$-0.31$ and $67.3\%$ for sunspot areas; and (4) $-0.33$ and $41\%$ for
$10.7$\,cm radio flux.  The results for sunspot numbers and sunspot
areas are shown in panels (a) and (b) of Fig.~1. It may be noted that
the data points of cycles 21, 22 and 23 which were largely
responsible for destroying the Waldmeier effect in
the analysis of Dikpati, Gilman \& de Toma (2008) are 
indicated by stars in Fig.~1(b) and are now quite close to the linear line.
These results are very sensitive to the averaging bin size. If we
average the data with a FWHM of $2$ yr instead of 1 yr, we obtain the following
correlation coefficients and significance levels for the four data
sets: (1) $-0.63$ and $97.28\%$; (2) $-0.60$ and $96.13\%$; (3) $-0.36$
and $75.38\%$; (4) $-0.67$ and $78.34\%$.  If we calculate the rise time
differently by taking the beginning and the end of the rise phase 
somewhat different from $0.2P$ and $0.8P$ (and also vary the FWHM
while averaging the data), then we get somewhat but not significantly
different correlation coefficients which are listed in Table~1 for 
sunspot area data.

\begin{table}
\caption[]{Linear correlation coefficients ($r$) and the significance
levels ($s.l.$) between the rise time and the peak value of sunspot area data.}
  \begin{center}\begin{tabular}{lrrrrr}
 \hline
                  & FWHM = 1 yr    & FWHM = 2 yr \\
\hline
 Rise time        & $r~(\%s.l.)$   & $r~(\%s.l.)$ \\
\hline
  $0.2P$ to $0.8~P $  & $-0.31~(67.3)$ & $-0.36~(75.4)$\\
  $0.2P$ to $0.7~P $  & $-0.40~(79.9)$ & $-0.42~(83.0)$\\
  $0.2P$ to $0.65P $  & $-0.43~(83.7)$ & $-0.35~(74.2)$\\
  $0.2P$ to $0.6~P $  & $-0.47~(87.6)$ & $-0.43~(83.7)$\\
  $0.15P$ to $0.8~P$ & $-0.32~(69.5)$  & $-0.40~(79.8)$\\
  $0.15P$ to $0.7~P$ & $-0.43~(83.3)$  & $-0.46~(86.7)$\\
  $0.15P$ to $0.65P$ & $-0.46~(86.7)$  & $-0.40~(80.4)$\\
  $0.15P$ to $0.6P $ & $-0.50~(90.0)$  & $-0.48~(88.2)$\\
\hline
\end{tabular}
  \end{center}
\end{table}

We also study the second Waldmeier effect WE2 in all four data
sets. We calculate the rise rate by determining the slope between
two points at a separation of one year, with the first point one year
after the sunspot minimum. We find strong correlation between the
rates of rise and the amplitudes of the sunspot cycles.  Results for
sunspot number and sunspot area are shown in panels (c) and (d) of
Fig.~1.   Cameron \& Sch\"ussler (2008) have
computed the rise rate slightly differently and obtained almost similar
results.

We conclude that there is evidence for both WE1 and WE2
in different kinds of data sets.

\subsection{Variations in meridional circulation}

Only from mid-1990s we have reliable data on the variation of meridional circulation.
Hathaway \& Rightmire (2010) analyze these data to conclude that the meridional
circulation varies with the sunspot cycle, becoming weaker at the time of the
sunspot maximum.  We should probably have to wait for at least one more full cycle
to reach a firm conclusion whether this variation indeed has the same period as the
sunspot cycle.  A systematic variation of the meridional circulation having
the same period as the sunspot cycle is not expected to introduce any irregularities
in the theoretically computed sunspot cycles.  Most of the dynamo calculations
presented by our group (Chatterjee, Nandy \& Choudhuri 2004; Choudhuri, Chatterjee, \& Nandy 2004;
Choudhuri, Chatterjee \& Jiang 2007; Jiang, Chatterjee \& Choudhuri 2007; Goel \& Choudhuri 2009; Karak \&
Choudhuri 2009) assumed a constant meridional circulation because even a few
years ago the available information about meridional circulation variation was very
scanty.

Since a periodic variation of the meridional circulation with the sunspot cycle
will not cause cycle irregularities, let us consider possible variations with
longer time scales which may affect sunspot cycles.  We have no direct information
about variations of meridional circulation prior to 1995.  However, if we believe
that the solar dynamo is a flux transport dynamo and the period of a cycle is
approximately inversely proportional to the meridional circulation during that
cycle, then we can draw some conclusions about the variations in meridional
circulation in the past from the periods of past sunspot cycles.  At the outset,
we point out that this is an unreliable and questionable procedure.  Since we
have no better way of inferring about variations in meridional circulation in the
past, it is still worthwhile to see what conclusions we can draw from this procedure.

\begin{figure*}
\centering
\includegraphics[width=1.00\textwidth]{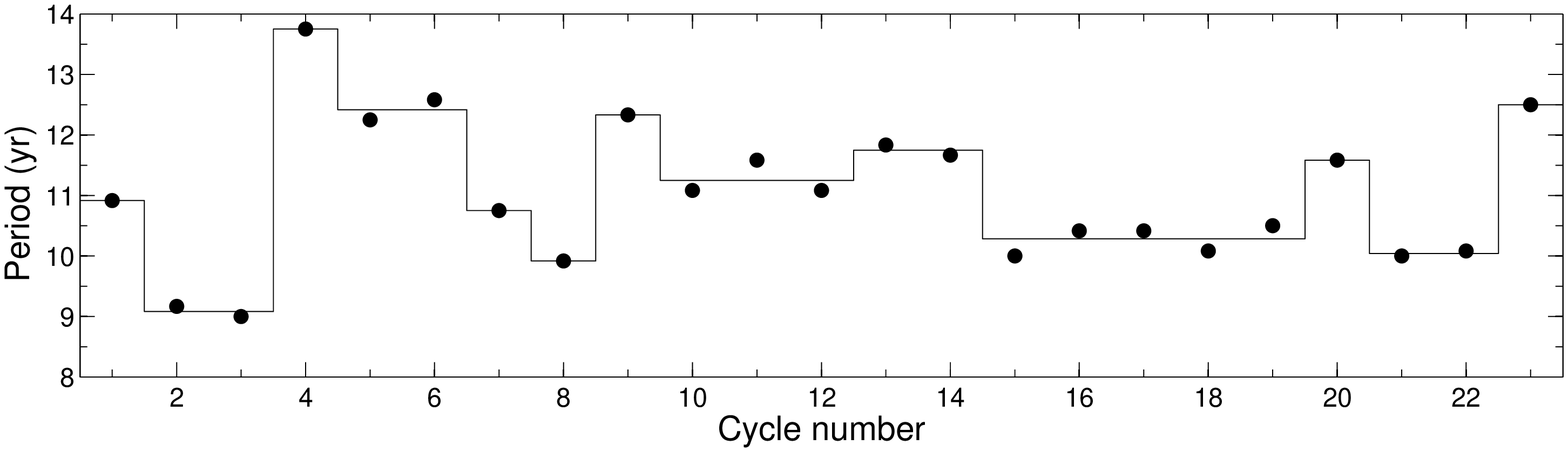}
\caption{The points show the periods of last 23 solar cycles against 
cycle number. The solid line is indicative of the trend in variations
of period as explained in the text.}
\label{period}
\end{figure*}

Fig.\ 2 shows the periods of the various sunspot cycles beginning with cycle~1.
If two successive cycles had similar periods, we may assume that the \mc\
had similar strengths during these two cycles. We have put a solid line in
Fig.~2 to indicate the trend of how periods of different cycles varied.
Whenever successive cycles had periods varying less than 5\% of the average 
period, we have made the solid line horizontal and indicative of the average period
of those cycles.  Only when periods of two successive cycles differed by
more than 5\%, there is a jump in the solid line in Fig.~2. If the solar
dynamo is a flux transport dynamo in which the period is set by the 
amplitude of \mc\, then the solid line should give an idea how the
\mc\ varied in the past. It appears that during cycles~1--10, the
\mc\ had a relatively short coherence time, but probably not less than 15 yr.
On the other hand, during cycles~11--20, the \mc\ seemed to have a longer coherence
time, but probably not longer than 45 yr. With the limited data we have, we cannot
say whether the behaviour of cycles~1--10 is more typical in the long run or
the behaviour of cycles~11--20 is more typical.   Looking at Fig.~2, we can
only surmise that the \mc\ probably has long-time variations having coherence
time lying somewhere between 15 yr and 45 yr.  In \S4 we shall present some
simulations assuming the coherence time of the \mc\ to be 30 yr. Since
meridional circulation and differential rotation both arise from turbulent
stresses in the convection zone, one intriguing and troubling question is
whether variations of \mc\ would be associated with the variations of differentia
rotation.  Since our theoretical understanding of this subject is still very
primitive and also to focus our attention on how variations of \mc\ affect
the dynamo, we have taken differential rotation to be constant in our paper.

To summarize, even though it is difficult to draw firm conclusions, it seems that
\mc\ has fluctuations having coherence times somewhat longer than a cycle---probably
of the order of 30 yr.  It may be noted that Charbonneau \& Dikpati (2000) argued
that the meridional circulation would have fluctuations with coherence time of the
order of a month, which is the eddy turnover time of solar convection.  We do
not find any observational signatures for such short-time fluctuations and such
short-time fluctuations of the \mc\ are not considered in this paper.

\section{Mathematical Formulation}

All our calculations are done with the code SURYA for solving the
axisymmetric kinematic dynamo problem.  An axisymmetric magnetic
field can be represented in the form
\begin{equation}
{\bf B} = B (r, \theta) {\bf e}_{\phi} + \nabla \times [ A
(r, \theta) {\bf e}_{\phi}],
\end{equation}
where $B (r, \theta)$ and $A(r, \theta)$ respectively correspond
to the toroidal and poloidal components.  The standard equations
for the kinematic dynamo are
\begin{equation}
\frac{\pa A}{\pa t} + \frac{1}{s}(\vf.\nabla)(s A)
= \eta_{p} \left( \nabla^2 - \frac{1}{s^2} \right) A + S(r, \theta, t)
\end{equation} 
\begin{eqnarray}
\frac{\pa B}{\pa t} 
+ \frac{1}{r} \left[ \frac{\pa}{\pa r}
(r v_r B) + \frac{\pa}{\pa \theta}(v_{\theta} B) \right]
= \eta_{t} \left( \nabla^2 - \frac{1}{s^2} \right) B \nonumber \\
+ s(\Bf_p.\nabla)\Omega + \frac{1}{r}\frac{d\eta_t}{dr}
\frac{\partial}{\partial{r}}(r B)
\end{eqnarray}\\
where $s = r \sin \theta$.  Here $\vf$ is the \mc, $\Omega$
is the internal angular velocity of the Sun, 
$S(r, \theta, t)$ is the source term for the poloidal field
by the Babcock--Leighton mechanism and $\eta_p$, $\eta_t$
are the turbulent diffusivities 
for the poloidal and toroidal components.

Since the internal rotation of the Sun has been determined by helioseismology,
most of recent dynamo models use a profile of the angular velocity $\Omega$
consistent with helioseismic findings.  Equation (8) of Chatterjee, Nandy \& Choudhuri
(2004) gives an analytical expression for $\Omega$ which is a good fit to
helioseismology results.  The profile of $\Omega$ obtained from this 
expression is shown in Fig.~1 of Chatterjee, Nandy \& Choudhuri (2004). We use this
$\Omega$ in all our calculations.  While the angular velocity is now
observationally constrained, different authors have modelled the meridional
circulation, the poloidal source term and the turbulent diffusivities
somewhat differently.  This has given rise to 
varieties of solar dynamo models.
In the last few years, however, two models have been studied fairly 
extensively---a high diffusivity model first developed by the group in 
Bangalore (Chatterjee, Nandy \& Choudhuri 2004)
and a low diffusivity model first developed by the group in Boulder
(Dikpati \& Charbonneau 1999). The turbulent diffusivities used in these models
are shown respectively in Fig.~4 of Chatterjee, Nandy \& Choudhuri (2004; solid line)
and Fig.~1(D) of Dikpati \& Charbonneau (1999).
Both these models use a fairly low
diffusivity in the tachocline (where turbulence is weak) to ensure
that the advection of the toroidal field by meridional circulation
dominates over diffusion there.  However, the turbulent diffusion within
the convection zone is assumed to be much larger.  What Chatterjee, 
Nandy \& Choudhuri (2004) call their `standard model' was produced with a diffusivity
of $2.4 \times 10^{12}$ cm$^2$ s$^{-1}$ for the poloidal field
within the convection zone, leading
to a diffusion time of a few years
across the convection zone. On the other hand, what Dikpati \&
Charbonneau (1999) call their `reference solution' was produced with a
much lower diffusivity of $5 \times 10^{10}$ cm$^2$ s$^{-1}$,
corresponding to a diffusion time of several centuries so that
the magnetic fields are virtually frozen during the period of
the dynamo. According to the classification scheme introduced by
Yeates, Nandy \& Mackay (2008), the model of Chatterjee, Nandy \& Choudhuri (2004) is
a `diffusion-dominated' model, whereas the model of Dikpati \&
Charbonneau (1999) is an `advection-dominated' model.

Jiang, Chatterjee \& Choudhuri (2007; \S5) gave several arguments that the diffusivity
within the convection zone is likely to be fairly high as assumed
by the group in Bangalore.  Subsequently several other authors also
have argued for high diffusivity (Goel \& Choudhuri 2009; Choudhuri
\& Karak 2009; Hotta \& Yokoyama 2010a, 2010b).  It appears that such a high
diffusivity is needed for getting the correct parity without an extra
poloidal source term within the convection zone (Chatterjee, Nandy \& Choudhuri 
2004; Hotta \& Yokoyama 2010b), for ensuring that the hemispheric asymmetry of magnetic
activity remains small as observed (Chatterjee \& Choudhuri 2006;
Goel \& Choudhuri 2009), for explaining the observed correlation of
the polar field with the strength of the next cycle (Jiang, Chatterjee \& Choudhuri
2007) and for keeping the polar field small in accordance with
observational data (Hotta \& Yokoyama 2010a). It may be noted that
straightforward mixing length arguments also suggest a high diffusivity,
Parker (1979, p.\ 629) concluding that the turbulent diffusivity
within the convection zone should be of order 
$1$--$4 \times 10^{12}$ cm$^2$ s$^{-1}$. We carry out our calculations
with the high diffusivity model and show that the theoretical model
predicts the Waldmeier effect roughly in accordance with the observational
data.  For the sake of completeness, we also explore the low diffusivity
Dikpati-Charbonneau (1999) model and find that this model is unable to 
explain WE1.

\begin{table}
\caption[]{The original values of the parameters in the standard
model (\S4 of Chatterjee, Nandy \& Choudhuri 2004) along with the changed values we
use here are given.}
  \begin{center}\begin{tabular}{c|c|c}
 \hline
Parameter & Standard Model & This Model \\
\hline\hline
$\eta_{SCZ}$ & $2.4\times10^{12}$ cm$^2$ s$^{-1}$ & $3.0\times10^{12}$ cm$^2$ s$^{-1}$ \\
$v_0$     & $-29$ m s$^{-1}$ &  $-23$ m s$^{-1}$ \\
$R_p$     & $0.61 R_{\odot}$ & $0.635 R_{\odot}$\\
$\alpha_0$& 25 m s$^{-1}$ & 30 m s$^{-1}$ \\
$\beta_2$ & $1.8\times10^{-8}$ m$^{-1}$ & $1.3\times10^{-8}$ m$^{-1}$ \\
$\Gamma$  & $3.47\times10^{8}$ m & $3.0253\times10^{8}$ m\\
$r_0$     & $0.1125 R_{\odot}$ & $0.1286 R_{\odot}$ \\
$d_{tac}$ & $0.025R_{\odot}$ & $0.03R_{\odot}$\\
\hline
\end{tabular}
  \end{center}
\end{table}

The `standard model' of Chatterjee, Nandy \& Choudhuri (2004) produced a
period somewhat larger than 11 yr. Also the value of the meridional
circulation near the surface was somewhat larger than what is observed.
For the calculations presented in this paper, we have changed some
parameters of the `standard model' to make the period 11 yr and to
make the meridional circulation at the surface equal to 23 m s$^{-1}$.
Table~2 lists the values of those parameters which have their
values changed in this paper from the values used by Chatterjee, Nandy \& Choudhuri 
(2004).  Except the values of these parameters listed in
Table~2, our model remains the same as the `standard model' of
Chatterjee, Nandy \& Choudhuri (2004).  We make a few comments on some aspects of 
this model. The meridional circulation in the
northern hemisphere is obtained from Equations (9)--(11) of Chatterjee, Nandy \& Choudhuri
(2004), from which we get the meridional circulation in the southern
hemisphere by antisymmetry.  Although we now choose some parameters of
the meridional circulation slightly differently from Chatterjee, Nandy \& Choudhuri (2004)
as listed in Table~2, the streamlines of meridional circulation still look
almost the same as in Fig.~2 of Chatterjee, Nandy \& Choudhuri (2004). The meridional
circulation used by us penetrates slightly below the bottom of the convection
zone, which is essential for confining the butterfly diagram to lower
latitudes (Nandy \& Choudhuri 2002). It may be noted
that there is a controversy at the present time whether the
meridional circulation can penetrate below the convection
zone---arguments having been advanced both against it (Gilman \& Miesch 2004) and
for it (Garaud \& Brummel 2008).  Recently Chakraborty, Choudhuri \& Chatterjee (2009)
have argued that the early initiation of torsional oscillations at latitudes
higher than the typical sunspot latitudes is possible only with such a
penetrating meridional circulation, providing another strong support for it.
The diffusion coefficients $\eta_p$ and $\eta_t$ are shown in Fig.~4 of
Chatterjee, Nandy \& Choudhuri (2004), where the justification for using two different
diffusivities is discussed.  Basically diffusion of the toroidal field is
suppressed inside concentrated flux tubes.  Since these flux tubes are
not resolved in the mean field theory, we capture this effect approximately
by making $\eta_t$ smaller than $\eta_p$ in the mean field equations.

For checking whether a low diffusivity model can explain the Waldmeier
effect, we have used the model of Dikpati \& Charbonneau (1999).  
It may be noted that this model, which produces butterfly diagrams
extending to high latitudes, was subsequently modified by Dikpati
et al.\ (2004) to build what they call a `calibrated flux transport
dynamo'.  It is this `calibrated flux transport dynamo' model which
was used by Dikpati \& Gilman (2006) to predict that the cycle~24 will be very
strong.  However, this `calibrated flux transport dynamo' of Dikpati
et al.\ (2004) has so far not been reproduced by any independent code
of any other group.  Some of the other groups who tried to reproduce
this model were unable to do so (Jiang, Chatterjee \& Choudhuri 2007; Hotta \& Yokoyama
2010a).  We also have tried to reproduce the results of Dikpati et al.\
(2004) and could not, although we are able to reproduce the results of
Dikpati \& Charbonneau (1999).  Jiang, Chatterjee \& Choudhuri (2007) noted that the
`reference solution' of Dikpati \& Charbonneau (1999) was reproduced
when the amplitude of meridional circulation was taken $u_0 = 20$ m s$^{-1}$
rather than $u_0 = 10$ m s$^{-1}$ as reported by Dikpati \& Charbonneau
(1999).  We also confirm this.  We have, however, taken the value
$u_0 = $ 14.5 m~s$^{-1}$ to ensure that the dynamo period comes out to
be 11 yr.  Everything else in the low diffusivity model we use in this
paper remains the same as in the `reference solution' of Dikpati
\& Charbonneau (1999).

To study whether the theoretical models can explain the Waldmeier
effect, we have to introduce irregularities in the theoretical model
to make the cycles unequal.  In the next section, we describe
how we introduce fluctuations in the poloidal field source term and
in the meridional circulation, and we present the results we get by
introducing these fluctuations.  To look for the Waldmeier effect,
we need to find out how the sunspot number varies with time. 
Charbonneau \& Dikpati (2000) proposed that the magnetic energy
density at latitude 15$^{\circ}$ at the base of the convection 
zone ($r = 0.7 \Rs$) can be taken to be a good proxy of the sunspot
number and used this to produce the sunspot number plots which they
presented.  We also take this as a proxy for the sunspot number in
this paper for both the high diffusivity and low diffusivity models.   

\section{Results from theoretical models}

We now present the results obtained by using both the high diffusivity
(or diffusion-dominated) model and the low diffusivity (or advection-dominated)
model introduced in \S3. After introducing irregularities
in the models, we generate the sunspot number plot for a particular
run by using the magnetic
energy density at latitude 15$^{\circ}$ at the base of the convection 
zone as the proxy of the sunspot number.  Then the rise time 
and the rise rate are calculated
exactly the way they were done for the observational data as described
in \S2.1.  We shall first present the results obtained by introducing
fluctuations in the poloidal field generation and then present results
with fluctuations in the \mc. It may be noted that Charbonneau \& Dikpati
(2000) presented some results by introducing these two kinds of fluctuations
in their low diffusivity model.  However, we introduce the fluctuations
somewhat differently and, in one important case, we find a result which
is opposite of what Charbonneau \& Dikpati (2000) presented, as we shall
point out.

\subsection{Fluctuations in the poloidal field generation}

As argued by Choudhuri, Chatterjee \& Jiang (2007) and Jiang, Chatterjee \& Choudhuri (2007), the
cumulative effect of fluctuations in the Babcock--Leighton process
which produces the poloidal field can be incorporated by stopping
the dynamo code at every minimum and then multiplying the poloidal
field above $0.8 \Rs$ by a factor $\gamma$.  We now present results
of runs for both the high and low diffusivity models in which $\gamma$
at a minimum was taken to be a random number lying in the range 0.5--1.5.

\begin{figure*}
\centering
\includegraphics[width=1.0\textwidth]{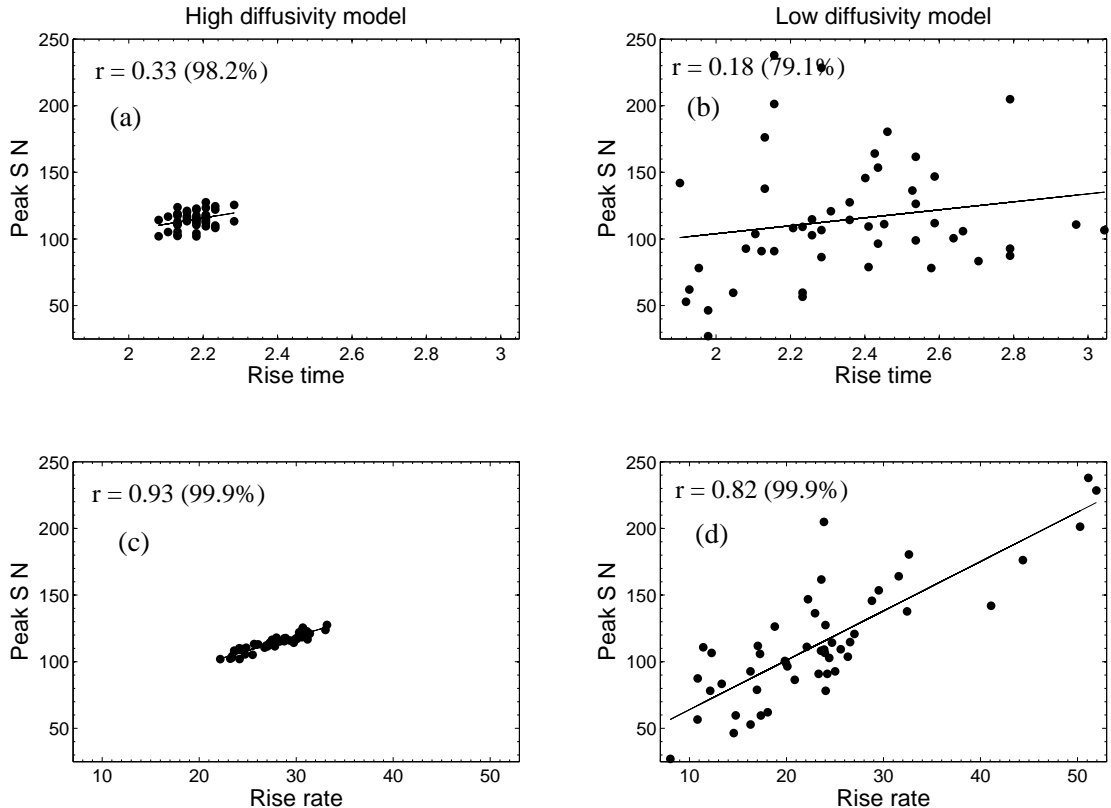}
\caption{Theoretical plots of WE1 (upper row) and WE2 (lower row) obtained 
by introducing fluctuations in the poloidal field at the minima.
The upper figures (a) and (b) show the scatter diagrams between 
the rise time (in years) and the peak sunspot number, 
whereas the lower figures (c) and (d) show the scatter diagrams between the 
rise rate (in units of sunspot number per year) and the peak sunspot number. The left figures 
(a) and (c) are from the high diffusivity model, whereas the right figures (b) and (d) 
are from the low diffusivity model. The straight lines are the best linear fits of the data. 
The correlation coefficients and the significance levels are also given on each plot.}
\label{pol}
\end{figure*}

\begin{figure*}
\centering
\includegraphics[width=1.0\textwidth]{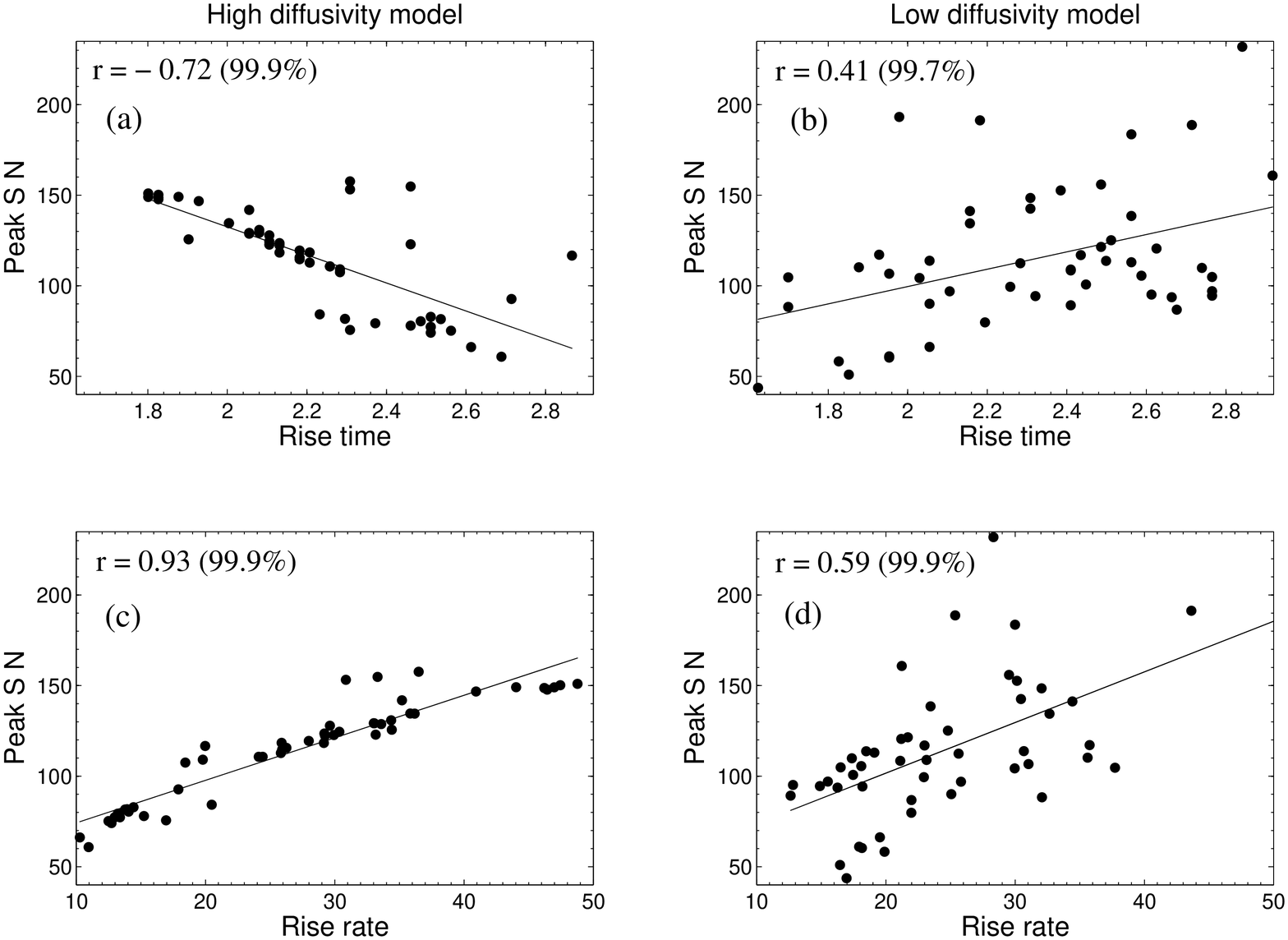}
\caption{Theoretical plots obtained by introducing fluctuations 
in the meridional circulation. The four figures correspond to the
same things as in Fig.~3.}
\label{mc}
\end{figure*}

\begin{figure*}
\centering
\includegraphics[width=1.0\textwidth]{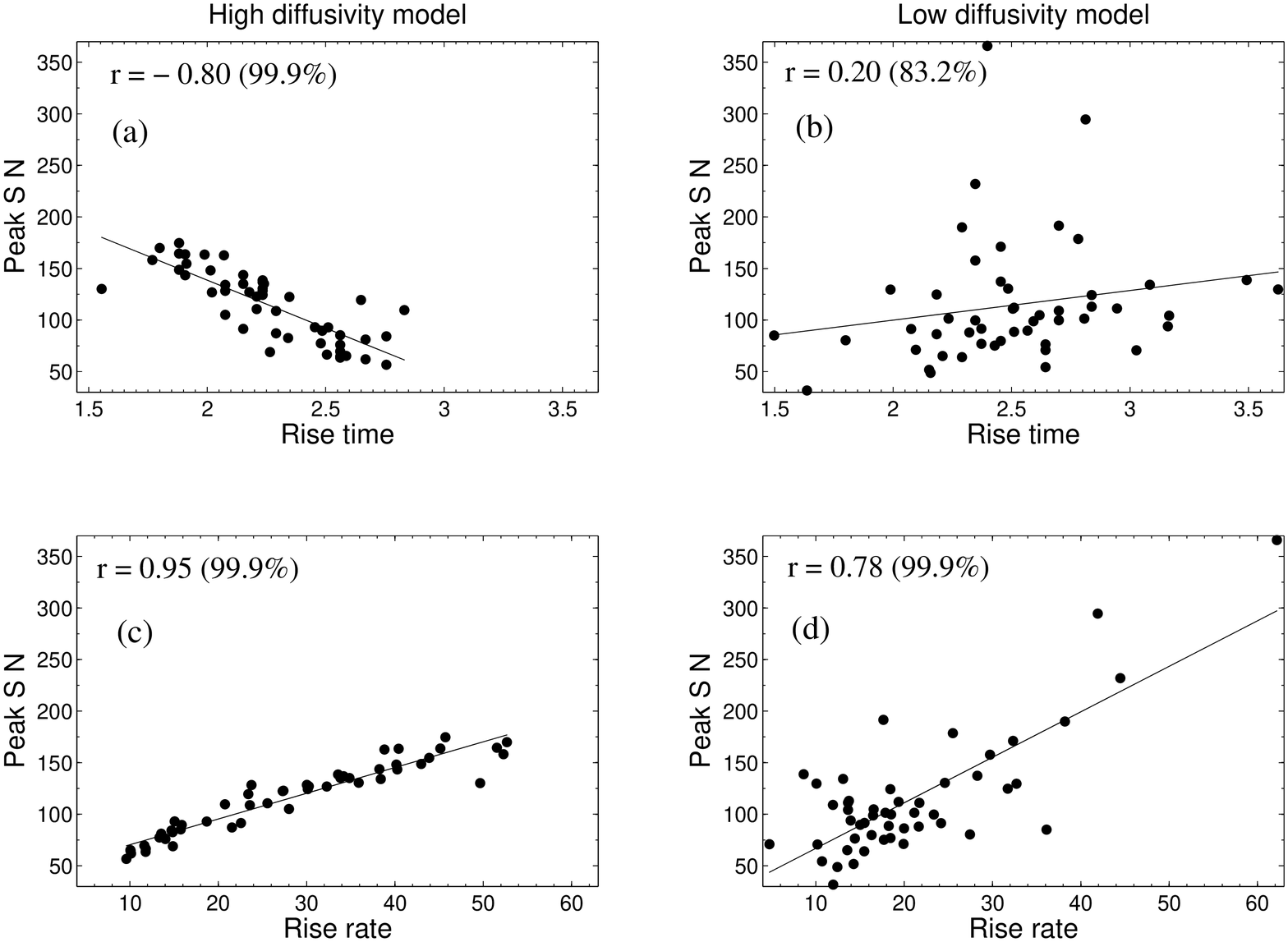}
\caption{Theoretical plots obtained by introducing fluctuations
in both the poloidal field generation and 
the meridional circulation. The four figures correspond to the
same things as in Fig.~3.}
\label{both}
\end{figure*}

First, let us look at the lower panels of Fig.~3 which show the correlations between the rise 
rate and the peak sunspot
number in the high and low diffusivity cases respectively.  In both the cases,
it is found that cycles with stronger peaks tend to have higher rise
rates, in accordance with the observed WE2.  It is easy to understand
why this is so.  As long as the \mc\ is held constant, the periods of
various cycles in a flux transport dynamo do not vary too much.  However,
fluctuations in the poloidal field generation make the strengths of
different cycles unequal.  A strong cycle has to rise to a higher
value of peak sunspot number compared to a weak cycle in approximately
the same amount of time. Therefore, a stronger cycle has to have a
higher rise rate. This is true irrespective of whether the diffusivity
is high or low.  We thus conclude that fluctuations in poloidal
field generation can easily account for the effect WE2. 

Although we believe that the effect WE1 involving the rise time
is primarily produced by fluctuations in the \mc, the upper panels
in Fig.~3 show the correlations between the rise time and the peak
sunspot number when fluctuations in the poloidal field generation
alone are present.  For both the high and low diffusivity cases, the
theoretical results (a positive correlation) are the opposite of 
the observational effect WE1 (anti-correlation).  In the high
diffusivity model, the fluctuations in the poloidal field generation
do not cause much variations in the cycle periods and the variations
in rise time are seen to be rather small in Fig.~3a.  On the other
hand, we see in Fig.~3b that rise times have a much larger spread
for the low diffusivity model.  Presumably fluctuations in the high
diffusivity model are damped out within  a few years and cannot
cause so much variations in the durations of cycles.  On the other
hand, fluctuations in the low diffusivity model persist for times
much longer than the period of the dynamo and can affect the
durations of cycles.  In both cases, however, fluctuations in
poloidal field generation alone cannot account for WE1.  We need
something else---presumably fluctuations in the \mc.

It may be noted that Charbonneau \& Dikpati (2000) reported a
weak anti-correlation between cycle duration and cycle amplitude
on introducing fluctuations in the poloidal source term 
(see their Fig.~6C). It is true that Charbonneau \& Dikpati (2000)
treated the fluctuations in the poloidal source term somewhat
differently from what we are doing and plotted the cycle
amplitude against the cycle duration rather than the rise time.
However, we repeated their procedure for the low diffusivity
model and found that we still get a weak correlation similar
to our Fig.~3b rather than the weak anti-correlation seen in
their Fig.~6C.  It should be noted that, when runs are repeated with 
different realizations of random numbers, the correlation
coefficients turn out to be somewhat different.
It is true that we and Charbonneau \& Dikpati (2000) 
got rather small correlation
coefficients of opposite sign: $r= 0.18$ and $r=-0.23$ respectively.
To some extent, these differences may be due
to statistical uncertainties in different numerical
realizations of the same problem involving fluctuations
created by random numbers.  However, in the several runs
we performed, we never got a negative correlation coefficient.
It will be worthwhile for other groups to check
this independently. 

The fact that fluctuations in the poloidal field generation do
not introduce much variations in cycle durations in the
diffusion-dominated model but introduce more variations in the
advection-dominated model has another significance.  The arguments
we have given in \S2.2 about variations in \mc\ are based on
the assumption that periods of cycles do not vary much as long
as the \mc\ is held constant.  As we now see, this is strictly
true only for the diffusion-dominated dynamo.  As we believe the
solar dynamo to be diffusion-dominated, the arguments we have given
in \S2.2 should be valid for the \mc\ in the Sun.   

\subsection{Fluctuations in the \mc}

We now study the results of introducing fluctuations in the
\mc.  As we argued in \S2.2, fluctuations in the \mc\ seem to have 
a coherence time of about 30 years if we believe that the periods of
past cycles were indicative of the variations in \mc.  We
run our code by changing the amplitude of the \mc\ abruptly
after every 30 years.  It appears that the high diffusivity
(or diffusion-dominated) model requires a stronger fluctuation 
in the \mc\ compared
to the low diffusivity (or advection-dominated) model to introduce the same kinds of
variations in cycle periods.  We use a 30\% amplitude fluctuation
in the high diffusivity model and a 20\% amplitude fluctuation
in the low diffusivity model.  

The results are shown in Fig.~4.  For both the models, the rise
rate is correlated with the peak sunspot number, as seen in the
lower panels of Fig.~4.  In other words,
the effect WE2 is reproduced from the theoretical models easily
whether the diffusivity is high or low.  However, we see a
dramatic difference between the two models when we look at
the plots of peak sunspot number against rise time (the two
upper panels in Fig.~4).  For the high diffusivity model, we
find that the rise time is anti-correlated with the peak sunspot
number, in accordance with the Waldmeier effect WE1.  On the
other hand, the low diffusivity model shows a correlation,
which is the opposite of the observed Waldmeier effect WE1.
We point out that Charbonneau \& Dikpati (2000) also reported
such a positive correlation (opposite of the Waldmeier effect)
on introducing fluctuations in the \mc\ in their low diffusivity
model (see their Fig.~4C), although they introduced the fluctuations
differently from what we are doing.

To understand the physics behind this dramatic difference between
the two models, the readers are advised to refer to Fig.~5 of
Yeates, Nandy \& Mackay (2008) and the associated discussion.  Let us
summarize the main argument. Suppose the \mc\ has become
weaker during a cycle.  Then the duration of the cycle will
be longer and the magnetic fields will spend
more time at the bottom of the convection zone.  This
will result in two opposing effects.  Diffusion will have
more time to act on the magnetic fields, trying to make the
cycle weaker.  On the other hand, differential rotation will
have more time to act on the poloidal field, building up a larger
toroidal field and making the cycle stronger.  Whether the
cycle will be weaker or stronger will depend on which of these
two effects win over.  In the high diffusivity (or diffusion-dominated)
model, diffusivity
acting on the magnetic fields is the more important effect.
Hence, when the \mc\ is weaker, the cycle duration (as well
as the rise time) is more and the strength of the cycle is
lower, leading to the anti-correlation in accordance with the
Waldmeier effect, as seen in Fig.~4a. 
The opposite of this happens in the low
diffusivity (or advection-dominated) model, where the differential 
rotation building
up a stronger toroidal field is the more important effect.
A weaker \mc\ causing a longer rise time will be associated
with a stronger cycle, opposite of the Waldmeier effect WE1,
as seen in Fig.~4b.

We thus see that the high and low diffusivity models give
very different results when fluctuations in the \mc\ are
introduced.  Only the high diffusivity model can explain the
Waldmeier effect WE1, while the low diffusivity model gives the
opposite result.  The effect WE2 is, however, explained by both
the models.

\subsection{Effect of combined fluctuations}

Finally we present results for cases where fluctuations in both
the poloidal field generation and the \mc\ are present.  As we
already mentioned, we have introduced fluctuations in the poloidal
field generation in \S4.1 in a way somewhat different from what
Charbonneau \& Dikpati (2000) had done.  In the calculations presented in
\S4.1, we have introduced the cumulative effect of poloidal
source fluctuations by multiplying the poloidal field above
$0.8 \Rs$ by a number $\gamma$ at each minimum.  We have also
done some calculations by introducing fluctuations in the poloidal
source term by the methodology of Charbonneau \& Dikpati (2000),
in which the amplitude of $\alpha$ is changed randomly after
a coherence time of 1 month, the level of fluctuations in the
amplitude of $\alpha$ being another parameter in the problem.
The results obtained by the two methodologies are found to be
very similar.  Here we now present results in which fluctuations
in \mc\ are introduced exactly as what we had done in \S4.2,
but fluctuations in poloidal field source are introduced by
the methodology of Charbonneau \& Dikpati (2000) which involves
a fluctuation in the $\alpha$-effect (Choudhuri 1992).  Both for
the high and low diffusivity models, the amplitude of $\alpha$
is changed after the coherence time 1 month, the level of fluctuations
being 100\% for the high diffusivity model and 200\% for the low
diffusivity model.  Charbonneau \& Dikpati (2000) had also used
200\% fluctuations in their low diffusivity model.

The results are shown in Fig.~5.  Since both kinds of fluctuations
taken individually produced a direct correlation between rise
rate and peak sunspot number in both the high and low diffusivity
models (the lower panels in Figs.~3 and 4), 
we naturally expect such a correlation to arise when
both kinds of fluctuations are combined.  This is clearly seen
in the lower panels of Fig.~5, indicating that WE2 is a robust
result and can be produced in theoretical models irrespective of
whether the diffusivity is high or low. However, when we look
for the Waldmeier effect WE1 in the correlation between rise time
and peak sunspot number, then the situation is more complicated.
For the low diffusivity model, both kinds of fluctuations taken
individually produced a direct correlation between them, which is
the opposite of the Waldmeier effect 
(Figs.~3b and 4b).  Not surprisingly, we see
a direct correlation when the two kinds of fluctuations are combined.
We thus conclude that the low diffusivity model cannot explain
the Waldmeier effect WE1.  For the high diffusivity model, we
had a direct correlation when fluctuations in poloidal field
generation alone were present (Fig.~3a) and an anti-correlation
when fluctuations in \mc\ alone were present (Fig.~4a), the spread
in rise times being rather small in Fig.~3a.  When both these kinds of
fluctuations are combined, we find an anti-correlation as seen
in Fig.~5a.  Thus the high diffusivity model reproduces the Waldmeier
effect WE1.  To sum up, the Waldmeier effect WE1 is reproduced
theoretically only in the high diffusivity model and not in the
low diffusivity model when fluctuations in both the poloidal
field generation and \mc\ are included.

It may be noted that, even in the high diffusivity model, we
need to make the coherence time of meridional circulation fluctuations
somewhat longer than the cycle period (we have taken 30 yr for
the results presented in Figs.~4--5) in order to obtain the
Waldmeier effect WE1.  If the coherence time is made comparable
to the cycle period (10 or 15 yr), then we do not get WE1.
For a coherence time of 20 yr for meridional circulation
fluctuations, we still find the Waldmeier effect WE1 with 
the correlation coefficient and the significance level 
equal to $-0.43$ and $99.8\%$ respectively for a particular 
run instead of $-0.80$ and $99.9\%$ indicated in Fig.~5a.

\section{Conclusion}

To the best of our knowledge, this is the first systematic effort
of addressing the question whether the Waldmeier effect can be
explained on the basis of flux transport dynamo models of the
sunspot cycle.  Along with the Waldmeier effect that rise times
of cycles are anti-correlated with cycle amplitudes, which we
call WE1, we also consider the related effect that rise rates
are correlated with cycle amplitudes, which we call WE2.  In
view of a recent controversy whether the Waldmeier effect exists
in different kinds of data, we point out that, if we define
rise times and rise rates carefully, then we find evidence for
both WE1 and WE2 in different kinds of data.

We can think of two main sources of irregularities in the
dynamo cycles: fluctuations in the Babcock--Leighton mechanism
and fluctuations in the \mc.  We study the effects of both
kinds of fluctuations on the dynamo models.  Since not much
is known about long-term fluctuations of the \mc, we analyze
the periods of the past sunspot cycles in \S2.2 to draw some
tentative conclusions about fluctuations of the \mc\ in the past.

The main conclusion of our paper is that the effects of fluctuations
are dramatically different in high and low diffusivity models.
This is not surprising.  Fluctuations in the high diffusivity model
damp out in a few years.  On the other hand, fluctuations in the
low diffusivity model take times longer than the dynamo cycle to
damp out.  The left panels in Figs.~3, 4 and 5 show results obtained
with the high diffusivity model, whereas the right panels show
results for the low diffusivity model.  Even a cursory look at
these figures shows that similar fluctuations produce much
larger dispersions in the low diffusivity model. As long as
the \mc\ is held constant, durations of cycles do not vary much
in the high diffusivity model even after introducing fluctuations
in the poloidal field generation process.  This is seen in Fig.~3a.
But this is not so true in the low diffusivity model, as
can be seen in Fig.~3b.

We find that the effect WE2 is very robust and is reproduced
easily in different types of dynamo models subjected to different
types of fluctuations, as seen in the bottom panels of Figs.~3,
4 and 5.  Basically, a stronger cycle rises to its higher peak
at a faster rate. The most important result of our paper is that
the Waldmeier effect WE1 arises from the fluctuations in the meridional
circulation and this happens only for the high diffusivity model.
The low diffusivity model gives the opposite result.  We pointed
out how we can understand this physically on the basis of the
analysis presented by Yeates, Nandy \& Mackay (2008).  In the high 
diffusivity (or diffusion-dominated)
model, the longer cycle allows the diffusivity to act for a
longer time and results in the cycle being weaker, in accordance
with the Waldmeier effect.  In the low diffusivity (or advection-dominated)
model, on the other hand, a longer cycle means that the differential
rotation builds up the toroidal field to a stronger value, thus
giving the opposite of the Waldmeier effect.  Jiang, Chatterjee \& Choudhuri
(2007, \S5) gave several arguments why the turbulent diffusivity
inside the convection zone has to be sufficiently high to ensure
that the diffusion time is not more than a few years. Several
subsequent authors reinforced this point (Goel \& Choudhuri 2009;
Choudhuri \& Karak 2009; Hotta \& Yokoyama 2010a, 2010b).  The fact that
only the high diffusivity model can explain the Waldmeier effect
makes the case still stronger that the solar dynamo is a high
diffusivity or diffusion-dominated dynamo.

We finally come to the last question whether the high diffusivity
model reproduces the observational data not only qualitatively,
but also quantitatively.  The unit of the sunspot number in the
theoretical plots is chosen in such a way that the sunspot number
of an average cycle comes out to be $114.5$ (which is the average value 
of the peak sunspot numbers of last $12$ cycles).  With this choice of
unit for the vertical axes in Figs.~3, 4 and 5, the theoretical
plots can be readily compared with the observational plots.
Perhaps Fig.~5 with both kinds of fluctuations present is the
appropriate figure to compare with observations.  We should
compare Fig.~1a with Fig.~5a and Fig.~1c with Fig.~5c.  Although
the theoretical plots have more data points than the observational
plots, a comparison of the values on the horizontal and vertical
axes shows that the spreads in rise rate, rise time and peak
sunspot number are comparable in the observational and theoretical
plots.  It is true that the observational plots seem to have a
little bit more scatter compared to the theoretical plots, which
is particularly evident when we compare Fig.~1a with Fig.~5a.
In spite of this, most readers will hopefully agree with us that
the comparisons between theory and observations seem reasonably
satisfactory, suggesting that the characteristics of the fluctuations we
had assumed in our theoretical analysis probably are not very
far from reality.  It should be kept in mind that such calculations
involving random numbers give slightly different results for
different runs.  The run which produced Fig.~5 was repeated
several times to ensure that the results for the different runs
were only slightly different. 

\medskip  
\noindent{\it Acknowledgments:}
We thank an anonymous referee for valuable comments, which helped in
improving the paper.
This work is partly supported by DST, India (project No.SR/S2/HEP--15/2007). 
BBK thanks CSIR, India for financial support.

\end{document}